\documentclass{ws-procs9x6}

\begin{document}

\title{SAID Results for GDH-Related Sum Rules}

\author{R.~A.~Arndt, W.~J.~Briscoe, I.~I.~Strakovsky,
and R.~L.~Workman}

\address{Center for Nuclear Studies, Department of Physics, \\
The George Washington University, Washington,  DC 20052, USA}

\maketitle

\abstracts{
Model-dependence of the single-pion contribution to the GDH, Baldin,
and forward spin-polarizability sum rules is explored, using the most recent
SAID multipole analysis. Results from SAID and MAID are compared.
}

We have calculated the GDH, Baldin and forward spin-polarizability 
($\gamma_0$) sum rules using our most recent analysis of single-pion
photoproduction data. This set of integrals is useful in revealing
discrepancies, as the integrands are weighted by different powers of
the photon energy. Values are tabulated below. Results are given
over the threshold region and full energy range to illustrate where
differences are most pronounced.  For the charged-pion contribution, 
this threshold component is comparable to the full result.

\begin{table}[ht]
\tbl{Comparison of the SM02~\protect\cite{sm02} and 
     recent MAID2000~\protect\cite{maid} calculations 
     for the GDH and Baldin integrals and the forward 
     spin polarizability from threshold to 2~GeV (for
     MAID to 1.25~GeV) [left set] and from threshold 
     to 200~MeV [right set] displayed as SAID/MAID.
     \vspace*{1pt}}
{\footnotesize
\begin{tabular}{|c|c|c|c|c|c|c|}
\hline
Reaction& \multicolumn{3}{c|}{to 2000~MeV}     & \multicolumn{3}{c|}{to 200~MeV}     \\[1ex]
{}      &   GDH   &  Baldin     &$\gamma_0$    &   GDH   &  Baldin     &$\gamma_0$   \\[1ex]
{}      &($\mu b$)&($10^-4fm^3$)&($10^-4fm^4$) &($\mu b$)&($10^-4fm^3$)&($10^-4fm^4$)\\[1ex]
\hline
{}      &{}       &{}           &{}            &{}       &{}           &{}            \\[-1.5ex]
$\pi^0p$&-142/-150&  4.7/4.7    &  -1.40/-1.47 &  -2/-1  &  0.1/0.1    &  -0.05/-0.04 \\[1ex]
$\pi^+n$& -45/-18 &  6.8/6.9    &   0.55/0.79  &  30/32  &  1.2/1.2    &   0.99/1.02  \\[1ex]
 proton &-187/-168& 11.5/11.6   &  -0.85/-0.68 &  28/31  &  1.3/1.3    &   0.94/0.98  \\[1ex]
\hline
$\pi^0n$&-148/-153&  4.6/4.6    &  -1.44/-1.50 &  -1/-1  &  0.1/0.1    &  -0.04/0.04  \\[1ex] 
$\pi^-p$&  11/33  &  8.3/8.8    &   1.36/1.64  &  42/47  &  1.7/1.8    &   1.39/1.53  \\[1ex]
neutron &-137/-120& 12.9/13.4   &  -0.08/0.14  &  41/46  &  1.8/1.9    &   1.35/1.49  \\[1ex]
\hline
\end{tabular}\label{tab1}}
\end{table}

\begin{figure}[t!]
\leftline{\psfig{file=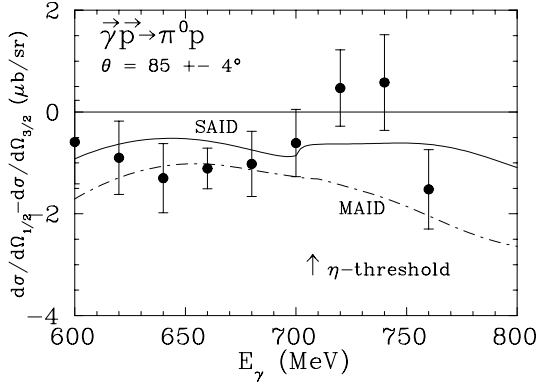,width=7cm,clip=,silent=,angle=90}}
\vspace{-5mm}
\hspace*{.65\textwidth}\raisebox{30mm}[0pt][0pt]
{\parbox{.35\textwidth}{\caption[fig1]{\label{fig1}Differential cross 
         section ($d \sigma /d \Omega_{1/2}-d \sigma /d \Omega
         _{3/2}$) for $\vec{\gamma}\vec{p}\to\pi^0p$ at $\theta 
         = 85\pm 4^{\circ}$.  The solid (dash-dotted) line plots 
         the SM02~\protect\cite{sm02} (MAID2001~\protect\cite{maid}) 
         solution.  Experimental data are from Mainz~
         \protect\cite{dx13}.}}}
\end{figure}

\begin{figure}[th]
\centerline{\psfig{file=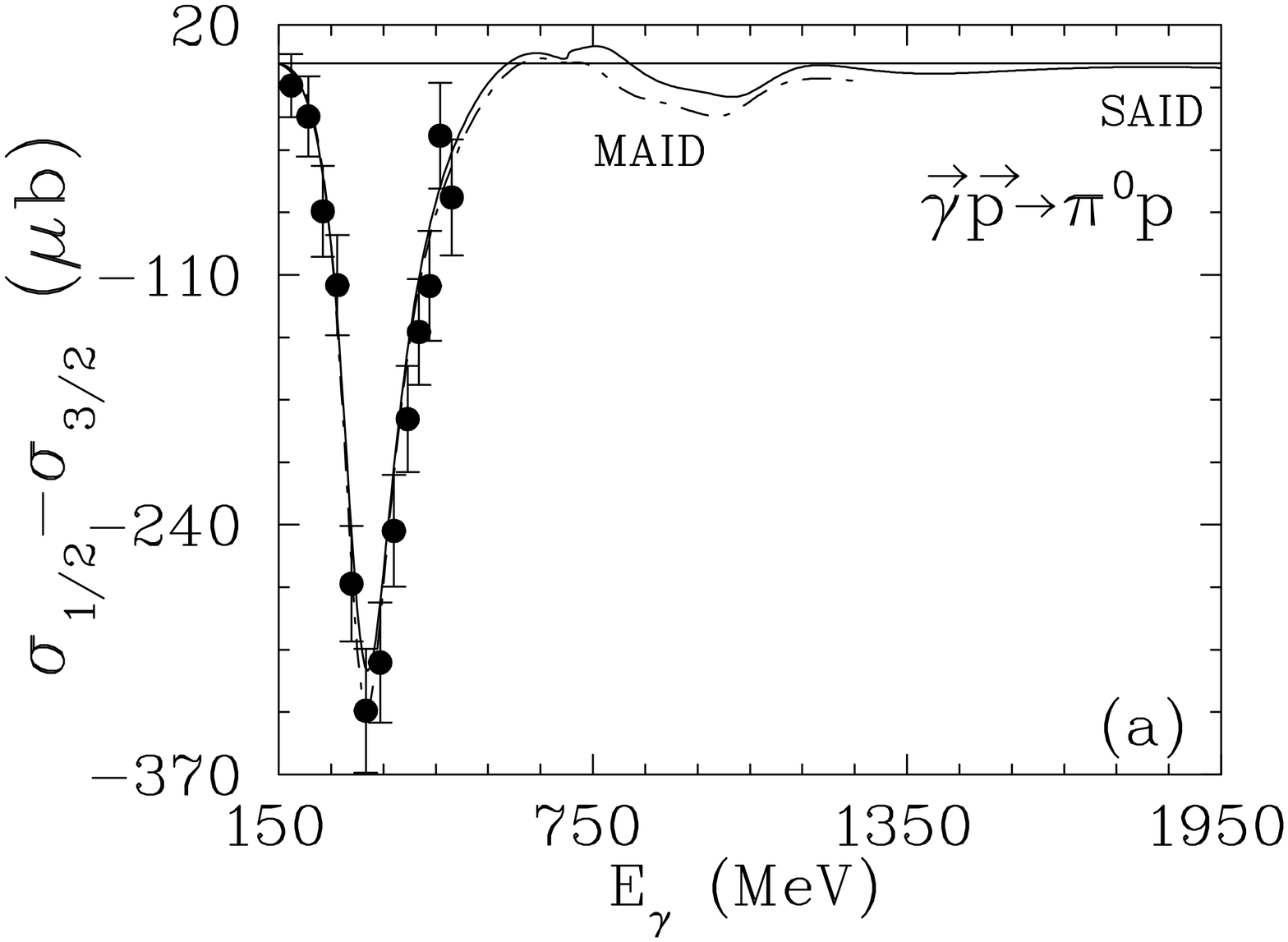,width=7cm,clip=,silent=,angle=90}\hfill
            \psfig{file=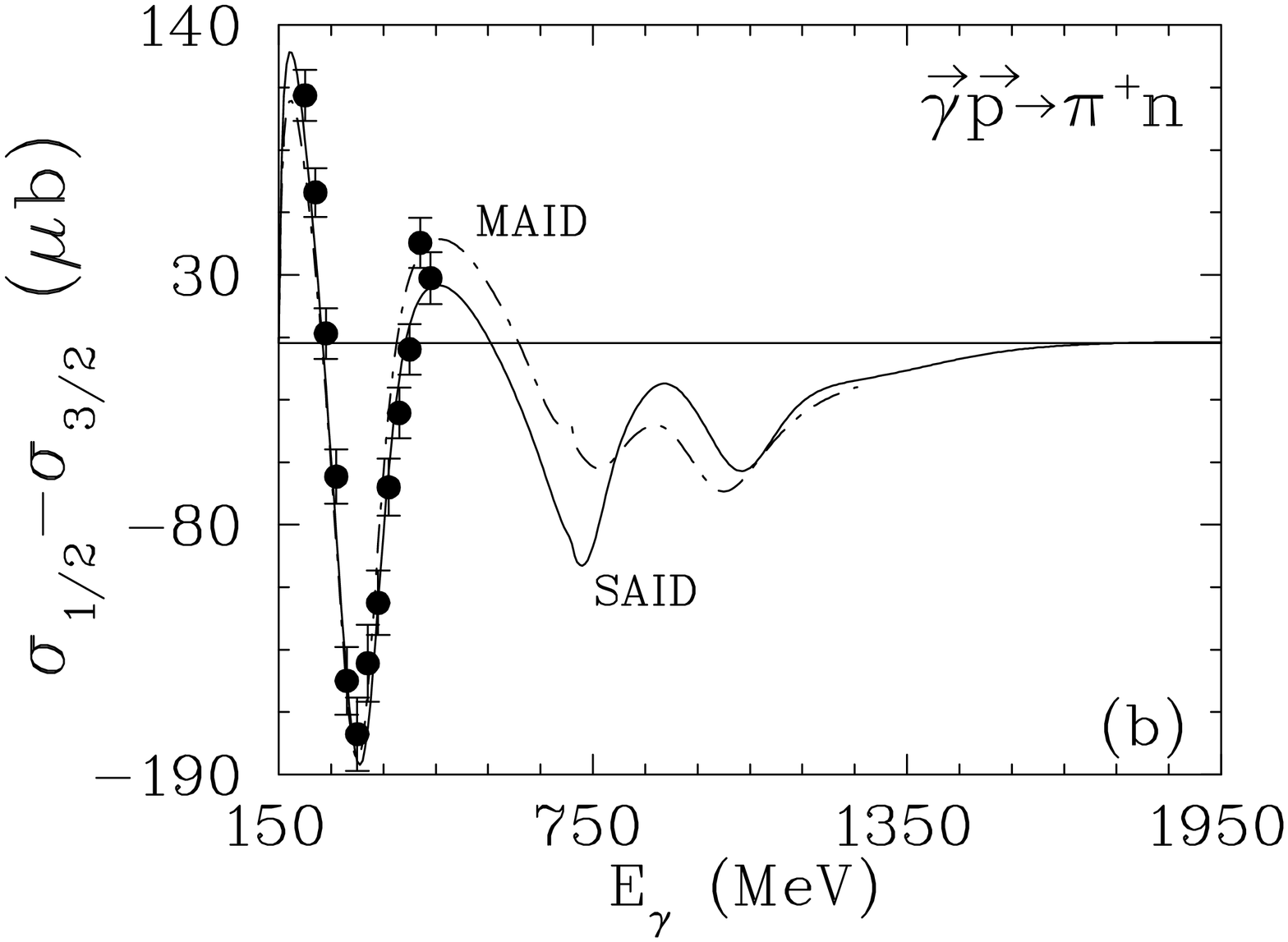,width=7cm,clip=,silent=,angle=90}}
\vspace*{8pt}
\caption{Difference of the total cross sections for the
         helicity states 1/2 and 3/2.
         (a) $\vec{\gamma}\vec{p}\to\pi^0p$ and
         (b) $\vec{\gamma}\vec{p}\to\pi^+n$.
         The solid (dash-dotted) line represents the
         SM02~\protect\cite{sm02} (MAID2001~
         \protect\cite{maid}) solution.
         Experimental data are from Mainz~
         \protect\cite{ah00}.\label{fig2}}
\end{figure}

\begin{figure}[th]
\centerline{\psfig{file=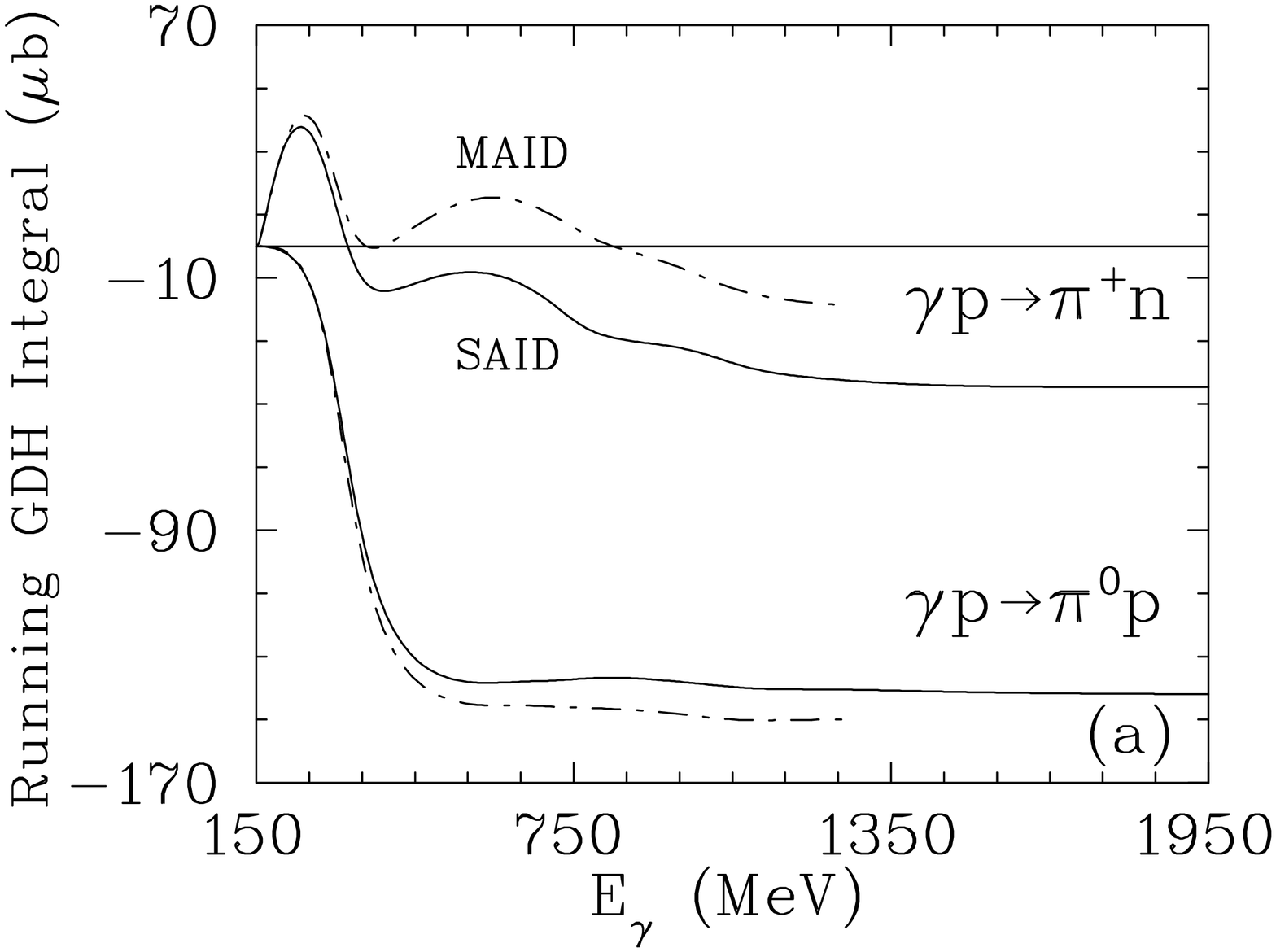,width=7cm,clip=,silent=,angle=90}\hfill
            \psfig{file=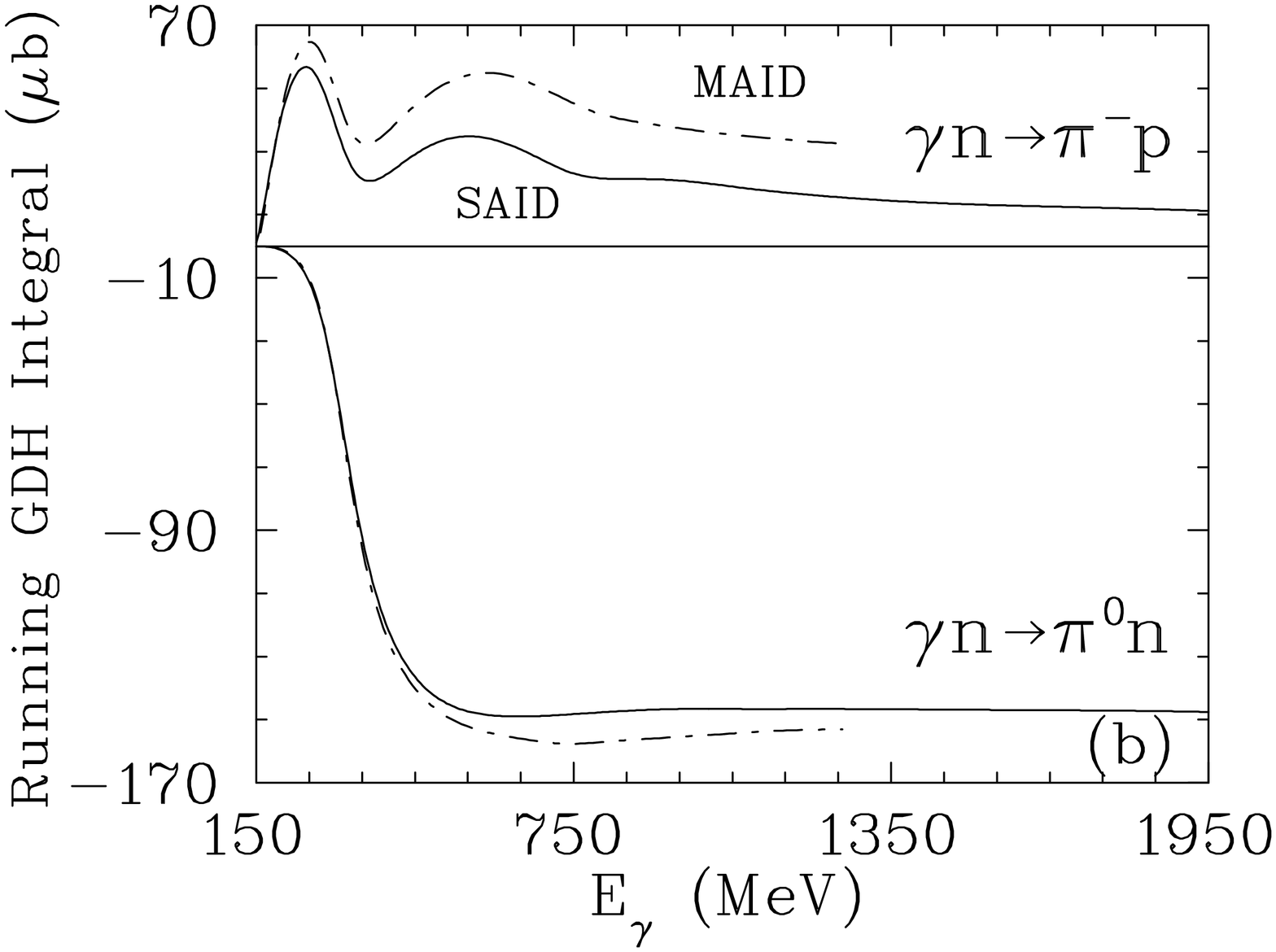,width=7cm,clip=,silent=,angle=90}}
\vspace*{8pt}
\caption{Running GDH integral.  (a) for proton and
         (b) neutron targets.  The solid (dash-dotted)
         line represents the SM02~\protect\cite{sm02} 
         (MAID2000
         ~\protect\cite{maid}) solution.\label{fig3}}
\end{figure}

\begin{figure}[th]
\centerline{\psfig{file=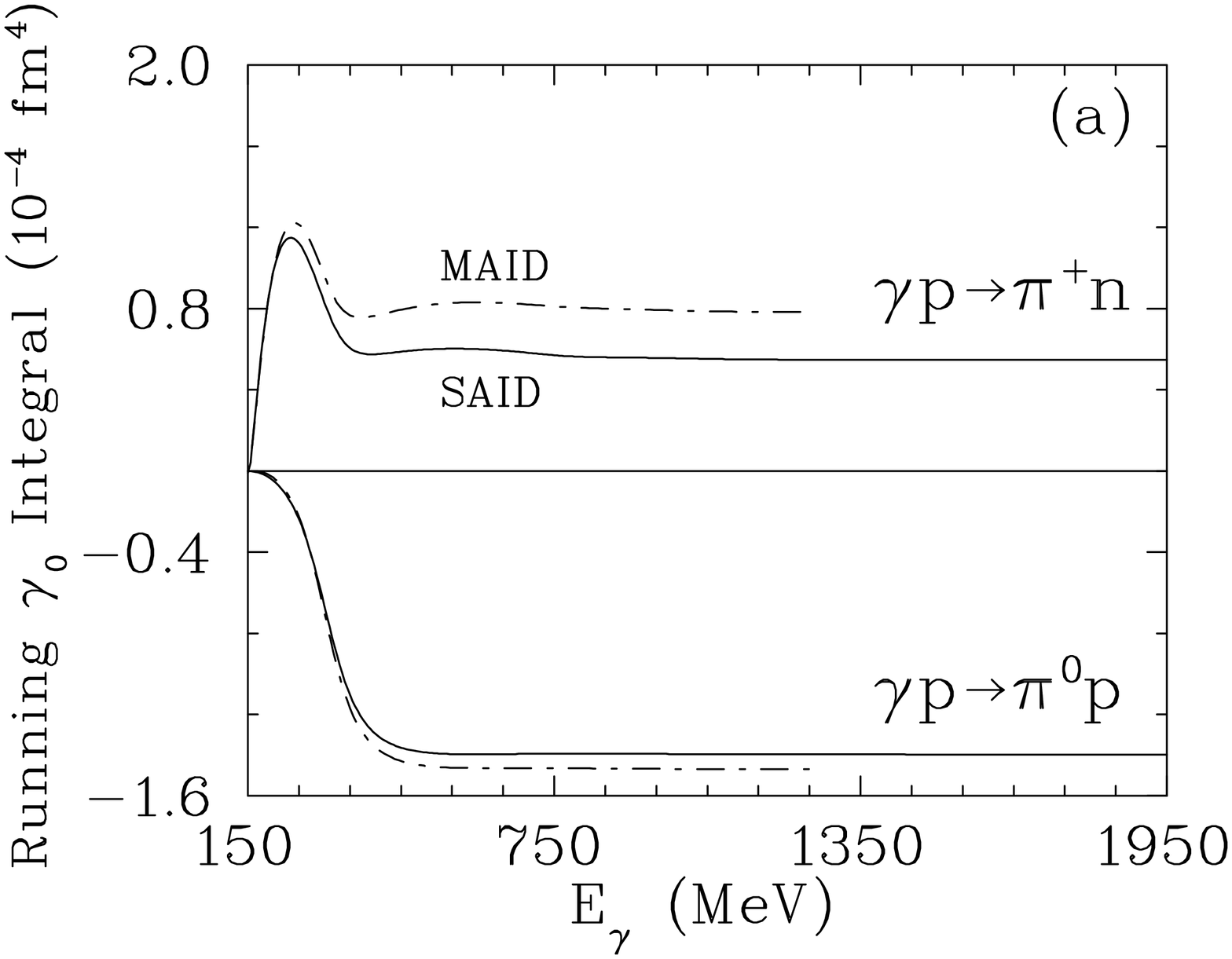,width=7cm,clip=,silent=,angle=90}\hfill
            \psfig{file=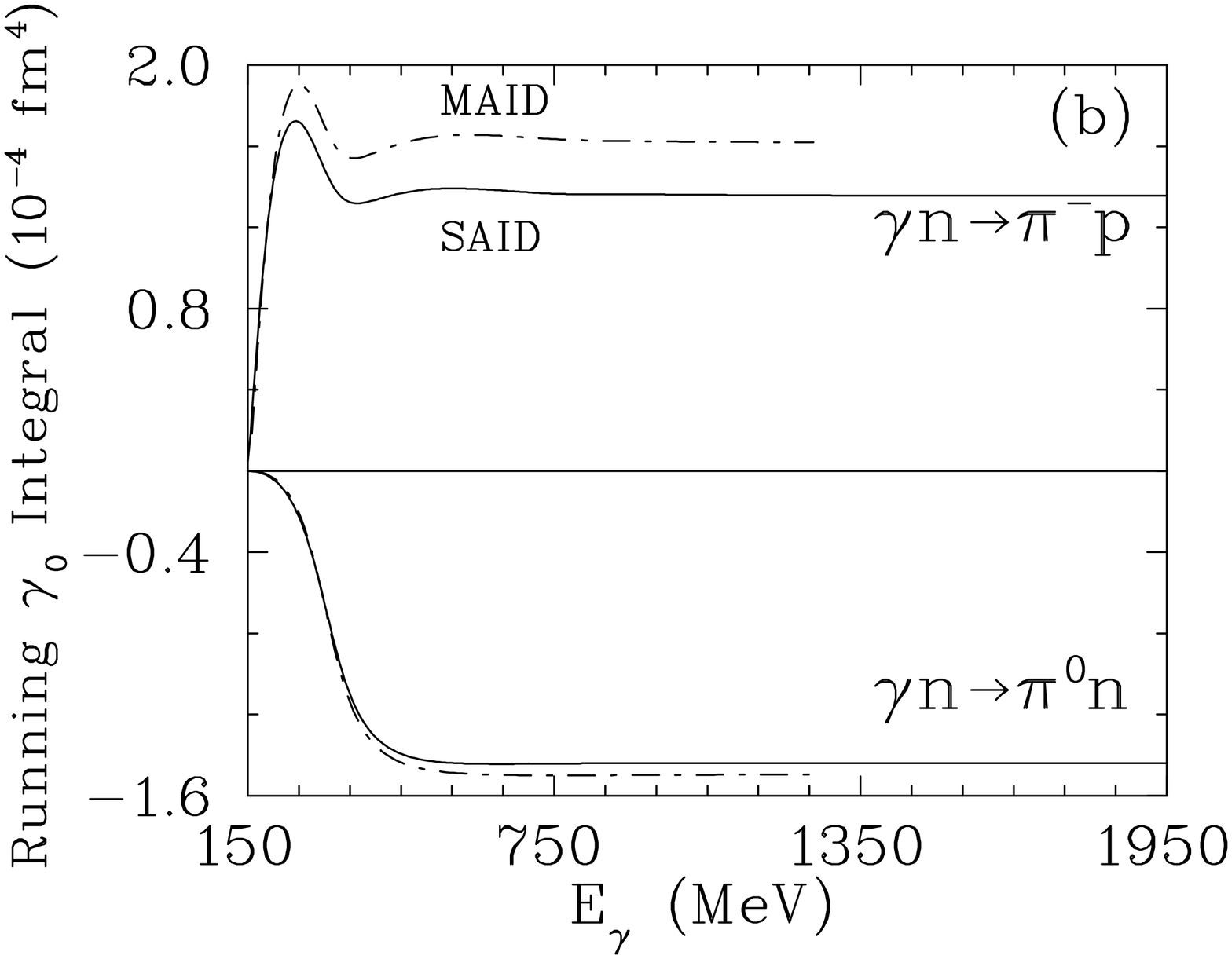,width=7cm,clip=,silent=,angle=90}}
\vspace*{8pt}
\caption{Forward spin polarizability $\gamma_0$.
         (a) for proton and (b) neutron targets.
         The solid (dash-dotted) line represents the
         SM02~\protect\cite{sm02} (MAID2000~
         \protect\cite{maid}) 
solution.\label{fig4}}
\end{figure}

Differences between SAID and MAID appear mainly above 450 MeV,
as illustrated in Fig.~\ref{fig1} (differential) and Fig.~\ref{fig2} 
(total) for the difference of helicity 1/2 and 3/2 cross sections. 
The running GDH and $\gamma_0$ integrals are given, for both proton 
and neutron targets, in Figures~\ref{fig3} and ~\ref{fig4}, respectively.  
For both integrals the single-pion contribution is negligible above 2~GeV. 
It is worth noting that the SAID and MAID results for total integrals
are more consistent than their individual pieces, due to cancellations 
which are evident in Fig.3.
Results for the Baldin integral are virtually identical for the 
$\gamma p$ components, with a slight discrepancy appearing for 
$\gamma n\to p\pi^-$. The good agreement for $\gamma n\to n\pi^0$ may
be misleading, with so little data existing in this channel. 


\begin{figure}[th]
\leftline{\psfig{file=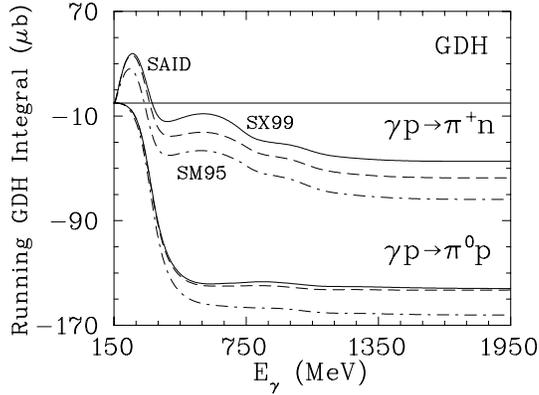,width=7cm,clip=,silent=,angle=90}}
\vspace{-5mm}
\hspace*{.65\textwidth}\raisebox{30mm}[0pt][0pt]
{\parbox{.35\textwidth}{\caption[fig5]{\label{fig5}Sensitivity 
         to parametrization/database.  The present fit
         (dashed) falls between SM95 and SX99.}}}
\end{figure}

In Fig.~~\ref{fig5} we compare GDH contributions calculated using an older
SAID fit (SM95), the present fit, and the result (SX99) of applying the
SM95 parametrization to the present database. From this it is clear
that both the changes in phenomenology and an improved database
have modified the integral. 

\vfil\eject

\section*{Acknowledgments}
This work has been supported by U.S. Department of Energy grant
DE-FG02-99-ER41110. Additional support has been provided by 
Jefferson Lab and the Southeastern Universities Research Association
through the D.O.E. contract DE-AC05-84ER40150.


\begin{thebibliography}{0}
\bibitem{sm02}R.~A.~Arndt, W.~J.~Briscoe, I.~I.~Strakovsky, 
              and R.~L.~Workman, submitted to {\it Phys.\ Rev.} 
              C., Eprint nucl-th/0205067, \\
              \hbox{http://gwdac.phys.gwu.edu}
\bibitem{maid}Mainz fits are available at the MAID website \\
              \hbox{http://www.kph.uni-mainz.de/MAID/}.
              See also S.~S.~Kamalov {\it et al.},
              {\it Phys.\ Rev.} C\ {\bf 64}, 032201 (2001).
              The Unitary Isobar Model is developed at
              Mainz (hereafter called MAID), D.~Drechsel
              {\it et al.}, {\it Nucl.\ Phys.} {\bf A645},
              145 (1999).  MAID2000 refers to the Feb.~2001
              version of the MAID solution, MAID2001 refers
              to the Nov.~2001 version of the MAID solution
              from S.~Kamalov.
\bibitem{dx13}J.~Ahrens {\it et al.}, {\it Phys.\ Rev.\ Lett.}
              {\bf 88}, 232002 (2002).
\bibitem{ah00}J.~Ahrens {\it et al.}, {\it Phys.\ Rev.\ Lett.}
              {\bf 84}, 5950 (2000).

\end{thebibliography}
\end{document}